\documentclass[12pt]{article}
\usepackage{amsmath}
\usepackage{amssymb}
\usepackage{amsthm}
\usepackage{psfrag}
\usepackage{graphicx}

\newcommand{\be}{\begin{equation}}
\newcommand{\ee}{\end{equation}}
\newcommand{\beqa}{\begin{eqnarray}}
\newcommand{\eeqa}{\end{eqnarray}}

\newcommand\m{\mu}

\renewcommand\a{\alpha}
\renewcommand\b{\beta}

\def\e{{\rm e}}
\def\d{\partial}
\newcommand{\bseq}{\begin{subequations}}
\newcommand{\eseq}{\end{subequations}}

\newcommand{\ch}{\mathop{\rm ch}\nolimits}
\newcommand{\sh}{\mathop{\rm sh}\nolimits}

\title{
\sc{\huge Supersymmetric Aether}
\date{}}
\author{Oriol Pujol\` as$^a$ and
  Sergey Sibiryakov$^{b}$\vspace{.2cm}\\
\normalsize\llap{$^a$}
 \it Departament de F\'isica and IFAE,
\normalsize\it Universitat Aut\` onoma de Barcelona,\\
\normalsize\it Bellatera 08193, Barcelona, Spain\\
\normalsize\llap{$^b$} \it Institute for Nuclear Research of the
Russian Academy of Sciences, \\ 
      \normalsize \it  60th October Anniversary Prospect, 7a, 117312
      Moscow, Russia
} 
\begin{document}

\maketitle

\begin{abstract}

It has been suggested by Groot Nibbelink and Pospelov that
Lorentz invariance can be an emergent symmetry of low-energy physics
provided the theory enjoys a non-relativistic version of supersymmetry.
We construct a model that realizes the latter symmetry dynamically: it breaks
Lorentz invariance but leaves the supersymmetry generators intact.
The model is a supersymmetric extension of the dynamical aether theory
of Jacobson and Mattingly.
It exhibits rich dynamics and possesses a family of inequivalent vacua realizing
different symmetry breaking patterns. 
In particular, we find stable vacua that break spontaneously spatial
isotropy.
Supersymmetry breaking terms give masses to fermionic and bosonic
partners of the aether field. 
We comment on the coupling of the model to supergravity and on the 
implications for Ho\v rava gravity.

\end{abstract}

\section{Introduction}

Recently the idea that Lorentz invariance (LI) may not be a
fundamental symmetry of nature has gained considerable attention. 
Indeed, violation of LI is present in one form or another in
many theories of gravity devised to solve the problems
of Einstein's
general relativity (GR). This is the case of the
infrared modifications of gravity, 
which attempt to address unresolved problems of cosmology such as
the nature of dark matter and dark energy. Examples are the ghost
condensate \cite{ArkaniHamed:2003uy}, Lorentz violating massive
gravity \cite{Dubovsky:2004sg}, the Galileons \cite{Nicolis:2008in} and
kinetic gravity braiding \cite{Deffayet:2010qz}. It has been
shown \cite{Blas:2011en} that violation of LI allows a simple and
technically natural explanation of the dark energy. Also breaking of
LI can be a consequence of quantum gravity. In particular, in the
approach to quantum gravity recently proposed by P.~Ho\v rava
\cite{Horava:2009uw} (see also
\cite{Blas:2009qj,Horava:2010zj,Blas:2010hb,Sotiriou:2011dr}) 
one manages to construct
a potentially renormalizable theory of quantum gravity at the price of
abandoning LI at very high 
energies.

Of course, LI is one of the best tested symmetries experimentally 
\cite{Mattingly:2005re,Liberati:2009pf,Kostelecky:2008ts} and one may
wonder how this can be reconciled with the ideas mentioned above. In
other words, in order to be viable, any Lorentz breaking model must
incorporate a mechanism that ensures the emergence of LI to a very
high accuracy, at least within the Standard Model (SM) of particle
physics at relatively low energies. 

In \cite{GrootNibbelink:2004za,Bolokhov:2005cj} it was suggested that
such mechanism can be provided by supersymmetry (SUSY). Namely,
consider the standard SUSY algebra and remove the boosts from
it \cite{Berger:2001rm,Berger:2003ay}
\footnote{
Refs.~\cite{GrootNibbelink:2004za,Bolokhov:2005cj,Berger:2001rm,Berger:2003ay}
  further restrict the SUSY algebra stripping off from it the spatial
  rotations as well. Here, we prefer to keep the invariance under
  rotations explicit and thus retain the corresponding generators in
  the symmetry algebra.}. The remaining generators still form a closed
algebra that we will call non-relativistic SUSY (or, where it will not
lead to confusion, just SUSY for short). As will be discussed below,
this is actually the most general minimal superalgebra
containing spatial rotations and space-time translations. A
theory invariant under the non-relativistic SUSY is in general not
Lorentz invariant: it is possible to explicitly construct
supersymmetric terms in the Lagrangian that violate LI. However, it
turns out that with the field content of the Minimal Supersymmetric
Extension of the SM (MSSM) all these Lorentz violating operators are of
dimensions five or higher\footnote{Dimension five operators can be
  forbidden by imposing additionally  CPT invariance, leaving
  dimension six terms as the lowest Lorentz violating
  contributions.}. In other words, LI emerges as an accidental
symmetry at the renormalizable level. The eventual breaking of SUSY
introduces Lorentz violation also at this level, but the effect is
within the existing bounds, provided the masses of the MSSM
superpartners lie sufficiently below the Lorentz violation scale
\cite{GrootNibbelink:2004za,Bolokhov:2005cj}. 

The analysis of \cite{GrootNibbelink:2004za,Bolokhov:2005cj} is
performed in the flat space-time where violation of LI corresponds to
the existence of a globally defined preferred frame. However, when
gravity is taken into account the space-time becomes dynamical and it
is clear that the preferred frame  must become dynamical as well. 
Thus to implement the mechanism of
\cite{GrootNibbelink:2004za,Bolokhov:2005cj} in gravitational models
with Lorentz violation it is necessary to construct SUSY extensions of
theories describing the preferred frame dynamics.

In the present paper we consider a representative theory of this class
-- the Einstein-aether model
\cite{Jacobson:2000xp,Jacobson:2008aj}. In this model the preferred
frame is defined by a  vector field $u^m$ (aether) which is
constrained to be time-like and have unit norm. 
We present a supersymmeric extension of
the model and analyze its 
consequences. We restrict to the case of global SUSY deferring
the coupling of the model to supergravity (SUGRA) for the future. 
This is a good approximation when the energy scale characterizing
Lorentz violation is small compared to the Planck mass.
Interestingly enough, we will find that 
the fluctuations of the aether field around its vacuum expectation
value (VEV) also exhibit LI at the level of low-dimension
operators. This extends the 
mechanism of \cite{GrootNibbelink:2004za,Bolokhov:2005cj} to the
supersymmetric aether sector.

The construction we present in this paper allows to realize 
the non-relativistic SUSY 
as the residual symmetry left over by the aether VEV in an otherwise 
super-Poincar\'e invariant setup. In this language the mechanism of 
\cite{GrootNibbelink:2004za,Bolokhov:2005cj} can be formulated as
follows: so long as the aether VEV preserves SUSY,
the Lorentz breaking does not propagate into the MSSM sector at the
renormalizable level. The eventual breaking of SUSY, that must be
incorporated in any realistic model, is unrelated to the dynamics of
the aether. It is assumed to come from a different source
characterized by a lower energy scale.
It is worth mentioning two intersting recent papers 
\cite{Khoury:2010gb,Khoury:2011da} where a general recipe was given to
construct supersymmetric versions of a wide class of theories
including Lorentz breaking models such as ghost condensation. 
However, in the models constructed in this way, the Lorentz violating 
backgrounds inevitably 
break SUSY at the same
time. This results in the absence of a hierarchy of scales between SUSY and Lorentz
breaking and so the LI of MSSM is not protected in these models. 
Our approach differs in spirit. Our main goal is to construct a
Lorentz violating theory with {\em unbroken} SUSY and study what kind
of restriction this requirement imposes on the structure of the
theory. 

The paper is organized as follows. In Sec.~\ref{aether} we describe the
Einstein-aether model. In Sec.~\ref{SUSY} we introduce 
the non-relativistic 
SUSY algebra and briefly review the arguments
of \cite{GrootNibbelink:2004za,Bolokhov:2005cj}. The SUSY extension of
the aether 
model is constructed in Sec.~\ref{superae}. In Sec.~\ref{breaking} 
we analyze the effects of SUSY breaking on the super-aether.
We conlude in Sec.~\ref{future} by
discussing our results and future directions.
\\

Throughout the paper we follow the conventions and notations of
\cite{WB}. In particular, the metric signature is $(-,+,+,+)$; Lorentz
indices are denoted by the letters from the middle of the latin
alphabet, $m,n,\dots = 0,1,2,3$; letters from the beginning of the
alphabet, $a,b,\ldots$, are used for purely spatial indices. We work
with two-component spinors, the spinor indices are denoted by Greek
letters with or without an overdot depending on the chirality. The
spinor indices are 
raised and lowered using the two-index antisymmetric tensor
$\varepsilon^{\a\b}$: 
$\psi^\a=\varepsilon^{\a\b}\psi_\b$,
$\psi_\a=\varepsilon_{\a\b}\psi^\b$,
where 
$\varepsilon^{12}=\varepsilon_{21}=1$.
We will often omit the spinor indices adopting the following
convention for the multiplication of spinors:
$\psi\eta \equiv\psi^\a\eta_\a$, 
$\bar\psi \bar\eta \equiv\bar\psi_{\dot\a}\bar\eta^{\dot\a}$.

\section{Review of Einstein-aether}
\label{aether}
The action of the Einstein-aether theory has the form
\cite{Jacobson:2000xp,Jacobson:2008aj}
\be
\label{Stotal}
S=S_{GR}+S_{\text{\ae}}\;,
\ee
where
\be
\label{SGR}
S_{GR}=-\frac{M^2_{pl}}{2}\int\sqrt{-g}\, R\,d^4x
\ee
is the standard Einstein-Hilbert action for gravity 
and the aether action is given by
\be
\label{Saether}
\begin{split}
S_{\text{\ae}}=-\frac{M^2_{\text{\ae}}}{2} \int\sqrt{-g}\Big(
c_1 \nabla_n u_m \nabla^n u^m
+c_2 (\nabla_m u^m)^2
&+c_3 \nabla_n u_m \nabla^m u^n\\
-c_4 \,u^r u^s \nabla_r u_m \nabla_s u^m
&+\lambda (u_m u^m+1)\Big)d^4x\;.
\end{split}
\ee
Here $u^m$ is a dynamical vector field and $\lambda$ is a Lagrange
multplier that enforces $u^m$ to be time-like with unit norm,
\be
\label{norm}
u_m u^m=-1\;.
\ee
The dimensionful parameter $M_{\text{\ae}}$ sets the scale of violation of LI;
the dimensionless constants $c_1,\ldots,c_4$ are the remaining free
parameters of the theory\footnote{This parameterization is redundant:
  the change of $M_{\text{\ae}}$ can be absorbed into redefinition of
  $c_i$'s. One could fix the ambiguity by choosing $M_{\text{\ae}}=M_{pl}$ as
  in \cite{Jacobson:2008aj}. Then the scale of Lorentz violation would
be set by the product $M_{pl}c_i$. However, we find convenient to keep
the scale of Lorentz breaking as an explicit parameter while assuming the
constants $c_i$ to be of order one.}. Eq.~(\ref{Saether}) is the most general
action for a vector with fixed norm containing up to two derivatives. 

Two comments are in order. The third term in (\ref{Saether})
can be cast after integration by parts into the same form as the
second term plus a non-minimal coupling of the vector to gravity,
\be
\label{equiv}
c_3\nabla_n u_m \nabla^m u^n\sim c_3(\nabla_m u^m)^2
+c_3 R_{mn}u^m u^n\;,
\ee
where $R_{mn}$ is the Ricci tensor. Next, the energy-momentum tensor
of the aether is proportional to $M_{\text{\ae}}$. Thus if the scale of
Lorentz breaking is much lower than the Planck scale, $M_{\text{\ae}}\ll
M_{pl}$, the effect of the aether on gravity is small. In this regime
it is self-consistent to consider the aether in external metric,
neglecting its back-reaction on the geometry. In this paper we
concentrate on the aether theory in flat space-time.

The ground state corresponds to a homogeneous
configuration of the aether which by a Lorentz transformataion can
always be cast into the form
\be
\label{vacuum}
u^m_{vac}=(1,0,0,0)\;.
\ee
This configuration breaks LI down to the $SO(3)$ subgroup
of spatial rotations. The model decsribes three propagating degrees of
freedom: one helicity-0 and two helicity-1 modes. These have
linear dispersion relations
\be
\label{disprel}
\omega=s_{(I)} |{\bf k}|
\ee
with the velocities\footnote{To obtain (\ref{va}) from the expressions of
  \cite{Jacobson:2008aj}
one should take the 
  limit $c_i\ll 1$
  corresponding to the case when back-reaction of the aether on 
gravity can be neglected.}
 \cite{Jacobson:2008aj}
\be
\label{va}
s_{(0)}^2=\frac{c_1+c_2+c_3}{c_1+c_4}~,~~~~~
s_{(1)}^2=\frac{c_1}{c_1+c_4}\;.
\ee
The difference in the propagation speeds of the modes with different
helicities clearly manifests the breakdown of LI.

Generally we can consider coupling the aether to other sectors of the
theory including SM. This will lead to a strong violation of LI in
these sectors. Let us illustrate this point on an example of a scalar
field $\phi$, that we take as a toy model for the SM
sector. Possible interactions of $\phi$ with the aether include a
dimension-four operator
\be
\label{LV4}
\frac{\kappa}{2}\, u^m u^n\, \d_m\phi\,\d_n\phi\;.
\ee
In the background (\ref{vacuum}) this will modify the dispersion
relations of the $\phi$-particles,
\be
\label{phidisp}
\omega^2=s^2{\bf k}^2+m^2\;,
\ee
where now the maximum propagation velocity $s$ is not equal to one,
\be
\label{phispeed}
s^2=\frac{1}{1+\kappa}\;.
\ee
For a single field this modification can be absorbed by redefinition
of the units of time or space. However, the real problem arises when
we consider several particle species. In general, their interactions
with the aether will involve different coupling constants leading to
different maximal propagation velocities. Non-observation of such
differences within SM leads to very tight constraints on the couplings
of the form (\ref{LV4}) \cite{Kostelecky:2008ts}. As we now discuss,
the unwanted couplings can be fobidden by SUSY.

\section{Lorentz invariance from non-relativistic SUSY}
\label{SUSY}

Consider the non-relativistic symmetry algebra consisting of the
generators of 3-dimensional translations $P_a$ and rotations $J_a$,
$a=1,2,3$, supplemented by the time translation $P_0$. Its minimal
supersymmetric extension contains two supercharges $Q_\a$, $\a=1,2$, 
transforming in the  spinor representation of the 
$SO(3)$ group of rotations, and their Hermitian conjugates
$\bar{Q}^\a$. The most general commutation relations compatible with
the $SO(3)$ symmetry and Jacobi identities are\footnote{We consider
  the case of non-zero $C$ in (\ref{3dSUSY3}). If $C=0$ there is an
  extra solution with 
\[
[P_0,Q_\a]=C_1 Q_\a+C_2\varepsilon_{\a\b}\bar{Q}^\b\;,
\]
where $C_1$ is real and $C_2=-C_1A/B$.}, 
\bseq
\label{3dSUSY}
\begin{align}
\label{3dSUSY1}
&\{Q_\a,Q_\b\}=2A \sigma_{\a\b}^a P_a\;,\\
\label{3dSUSY2}
&\{\bar Q^\a,\bar Q^\b\}=-2A^* (\sigma^a)^{\a\b} P_a\;,\\
\label{3dSUSY3}
&\{Q_\a,\bar Q^\b\}=2B (\sigma^a)_\a^{\b}P_a-2C\delta_\a^\b P_0\;,\\
\label{3dSUSY4}
&[P_a,Q_\a]=[P_a,\bar{Q}^\a]=[P_0,Q_\a]=[P_0,\bar{Q}^\a]=0\;.
\end{align}
\eseq
Here $A$ is a complex constant, $B$ and $C$ are real, and
$(\sigma^a)_\a^\b$ are the Pauli matrices\footnote{We limit our
  consideration to   {\em proper} algebras, in the sense that the
  r.h.s. of the commutation relations involves the generators only  linearly. 
  As was pointed out in
  \cite{Xue:2010ih,Redigolo:2011bv}, one could consider more general
  constructions allowing the coefficients in (\ref{3dSUSY}) to depend
  on the Casimir operators such as $P_a P_a$ and $P_0$. We do not
  pursue this route in the present paper.}. 
The commutators involving the angular momenta 
$J_a$ have the standard form dictated by the
$SO(3)$ symmetry and we do not write them explicitly. Note that by
choosing the dimension of the supercharges to be $(\mathrm{length})^{-1/2}$ the
constants $A$ and $B$ are made dimensionless. Then $C$ has the
dimension of velocity. Existence of this universal velocity encoded in
the SUSY algebra allows to connect units of space and time and, as we
are going to see, eventually leads to the emergence of LI at low energies.
We will refer to the algebra (\ref{3dSUSY}) as ``non-relativistic
SUSY''. 

We now
show that it is equivalent to the standard 4-dimensional
super-Poincar\'e algebra with 
the boosts stripped off. Indeed, it is straightforward to check that
by an appropriate linear transformation of the supercharges
\be
\label{Qredef}
Q_\a\mapsto \tilde Q_\a=a_1 Q_\a +a_2 \varepsilon_{\a\b}\bar Q^\b
\ee
the coefficient $A$ can be set to zero. 
Then, assuming the generic
case when both $B$ and $C$ are non-zero, 
we can set $B=C=1$ by
rescaling the generators $P_a$, $P_0$ 
(i.e., choosing appropriately the units of length and time). 
Finally, redefining
\be
\label{barQredef}
\bar{Q}^\a\mapsto\bar{Q}_{\dot\a}
\ee
one obtains
\bseq
\label{4dSUSY}
\begin{align}
\label{4dSUSY1}
&\{Q_\a,\bar{Q}_{\dot\beta}\}=2\sigma^m_{\a\dot\beta}P_m\;,\\
\label{4dSUSY2}
&\{Q_\a,Q_\b\}=\{\bar{Q}_{\dot\a},\bar{Q}_{\dot\b}\}=[P_m,Q_\a]
=[P_m,\bar{Q}_{\dot\a}]=0\;,
\end{align}
\eseq
where $\sigma^0_{\a\dot\b}\equiv-\delta_{\a}^{\dot\b}$. This is
nothing but the commutation relations of  4d SUSY.

We now briefly review the argument of
\cite{GrootNibbelink:2004za,Bolokhov:2005cj} showing that the symmetry
(\ref{4dSUSY}) leads to LI at the renormalizable level of MSSM. As in
the relativistic case, the representations of the algebra
(\ref{4dSUSY}) are classified in terms of the superfields: indeed, the
boost generators never appear in the superspace construction. The MSSM
Lagrangian is constructed out of a number of chiral matter
$\Phi_{(I)}$ and real gauge $V_{(J)}$ superfields. It involves
integration over the Grassmann variables $\theta_\a$: $\int
d^2\theta$ in the superpotential term and $\int d^2\theta
d^2\bar\theta$ in the kinetic (K\"ahler) part. Table \ref{tab:1} 
summarizes the mass dimensions of the objects that can appear in the
Lagrangian.
\begin{table}[htb]
\begin{center}
\begin{tabular}{|l|c|}
\hline
matter field $\Phi_{(I)}$ & 1\\ \hline
gauge field $V_{(J)}$& 0\\ \hline
gauge field strength $W_{(J)\,\a}$&3/2\\ \hline
space-time derivative $\d_m$&1\\\hline
supercovariant derivatives $D_\a$, $\bar{D}_{\dot\a}$&1/2\\\hline
chiral measure $\int d^2\theta$&1\\\hline
super-space measure $\int d^2\theta
d^2\bar\theta$& 2\\\hline
Lorentz breaking VEV $u^m$ & 0\\ \hline
\end{tabular}
\end{center}
\caption{The mass dimensions of various objects entering the MSSM
  Lagrangian.}
\label{tab:1}
\end{table}

Let us try to construct from these ingredients a Lorentz violating
contribution to the Lagrangian. 
This contribution must be invariant
under the remaining symmetries of the theory such as gauge
invariance.
It is convenient
to adopt a formally Lorentz covariant description using the vector
$u^m$ with a Lorentz breaking VEV (\ref{vacuum}) as a 
compensator\footnote{
In Sec \ref{sec:vacua} we will see that the SUSY aether can provide
richer patterns of Lorentz breaking. For the sake of the argument, we
concentrate here on the simplest pattern corresponding to the VEV
(\ref{vacuum}).}.
Then a
Lorentz violating term in the Lagrangian will contain a number of
Lorentz indices contracted with $u^m$. The superfields $\Phi_{(I)}$,
$V_{(J)}$ are Lorentz scalars. Thus the Lorentz violating term must
involve at least one derivative or the gauge field strength
$W_{(J)\,\a}$. By inspection, with the account of Table~\ref{tab:1},
one finds that there are no Lorentz violating terms of dimensions 2 or
3. The only allowed term of dimension 4, that is not a total
derivative, has the form
\be
\label{LVdim4}
{\cal L}_{LV}^{(4)}=\kappa_{(IJ)} \; u^m \int d^2\theta\; \Phi_{(I)}\d_m\Phi_{(J)}
+\mathrm{h.c.}\;,
\ee  
where $I\neq J$. However, in MSSM this term is forbidden by the gauge
invariance. Indeed, all fields of MSSM are charged under the gauge
group and the plain derivative in (\ref{LVdim4}) is not gauge
invariant\footnote{Note that terms of the form (\ref{LVdim4}) do
  appear if we include into consideration sterile neutrinos that are
  gauge singlets.}. One could try to remedy this by introducing the
gauge-covariant derivative 
\be
\label{covdiv}
{\cal D}_m\Phi=-\frac{i}{4}\bar\sigma_m^{\dot\a \a}\bar D_{\dot\a}
\e^{-V} D_\a\e^V\Phi\;.
\ee
However, this expression is no longer chiral,
\[
\bar D_{\dot\beta}{\cal D}_m\Phi\neq 0\;,
\]
and cannot be used in the superpotential term. 
Thus one concludes that in MSSM it is impossible to write any
Lorentz violating operator of dimension 4 or less.

To avoid confusion let us stress that the above argument does not
imply complete restoration of LI. One can easily construct higher
dimension operators that do violate LI. An example is the dimension 6
contribution 
\be
\label{LVdim6}
{\cal L}_{LV}^{(6)}=
\frac{u^m u^n}{M_{LV}^2}\int d^2\theta d^2\bar\theta\; 
{\cal D}_m\bar\Phi{\cal D}_n\Phi
\ee 
in the kinetic part. Here $M_{LV}$ is the scale of Lorentz
violation. Due to the eventual SUSY breaking the Lorentz violation
from higher-order operators feeds into the lower-dimensional ones 
\cite{GrootNibbelink:2004za,Bolokhov:2005cj}. In particular, the
operator (\ref{LVdim6}) gives rise to the dimension 4 terms of the form
(\ref{LV4}). However, the coefficients in front of these terms are
suppressed by the ratio $m^2_{SUSY}/M^2_{LV}$, where $m_{SUSY}$ is the
scale of SUSY breaking in MSSM (the scale of the superpartner
masses). This ratio can be small enough to satisfy the experimental
bounds on Lorentz violation.   

\section{Super-aether}
\label{superae}

\subsection{The Lagrangian}

We now turn to the construction of the supersymmetric aether
theory. We impose the requirement that this theory must be compatible
with the mechanism for emergence of LI reviewed in the previous
section. This determines the choice of the SUSY multiplet to embed the
aether vector $u^m$. Indeed, for the argument of the previous section
to hold it is necessary that a constant VEV
of $u^m$ preserves SUSY. This implies that $u^m$ must be the lowest
component of the multiplet. The simplest multiplet with these
properties corresponds to a chiral superfield $U^m$,
\be
\label{Uchiral}
\bar{D}_{\dot\a}U^m=0\;.
\ee
This superfield is a Lorentz vector. In components we write
\be
\label{Umu}
U^m=u^\m(y)+\sqrt{2}\theta^\a \eta^m_\a(y)
+\theta^2 G^m(y)\;,
\ee
where $\theta^2\equiv\theta^\a\theta_\a$ and 
$y^m=x^m+i\theta\sigma^m\bar\theta$. Note that the aether vector $u^m$
is now allowed to be complex,
\be
\label{ucomplex}
u^m=u_\mathrm{R}^m+iu_\mathrm{I}^m\;,
\ee
with $u_\mathrm{R}^m$, $u_\mathrm{I}^m$ real. The vector-spinor field
$\eta_\a^m$ in (\ref{Umu}) is the aether superpartner and $G^m$ is the
auxiliary component. By analogy with Eq.~(\ref{norm}) we impose the
constraint\footnote{The alternative choice of the constraint
  $U^m \bar{U}_m=-1$ does not lead to a consistent theory.}
\be
\label{Uconstr}
U^m U_m=-1\;.
\ee  
Note that $U^m$ has vanishing mass dimension.

Following the ordinary aether theory as the guideline 
we look for the super aether Lagrangian in the form 
\be
{\cal L}=M^2_{\text{\ae}}\tilde{\cal L}\;.
\ee
In other words, we factor out the scale of Lorentz breaking from the
Lagrangian. As long as we are interested in the low-energy physics we
can restrict to the operators of dimensions up to 2 in $\tilde{\cal
  L}$. We also assume that the aether field is the only source of
  Lorentz violation, and thus the Lagrangian must be a Lorentz scalar.
With these restrictions the only possible form of the kinetic
term is a function of the dimensionless  combination $U^m
\bar{U}_m$ because the superspace integration already contributes
dimension 2, see Table~\ref{tab:1}. Turning to the superpotential
part, all terms without derivatives are trivial due to the constraint
(\ref{Uconstr}). Moreover, terms with one space-time derivative vanish
due to the identity
\[
U^m\d_nU_m=0
\]
that is a consequence of (\ref{Uconstr}). Finally, the terms with more
derivatives are of higher dimensions. One concludes that at the level
of operators of dimensions less or equal 2 the superpotential
vanishes. In this way we arrive at the most general super-aether
Lagrangian 
\be
\label{SUSYaether}
{\cal L}^{(2)}=M^2_{\text{\ae}}\bigg[\int d^2\theta d^2\bar\theta \; f(U^m \bar{U}_m)
+\bigg(\int d^2\theta\;\Lambda(U^m U_m+1)+\mathrm{h.c.}\bigg)\bigg]\;,
\ee 
where $f$ is an arbitrary function and we have 
implemented the constraint (\ref{Uconstr}) by a superpotential term
with a Lagrange multiplier chiral superfield $\Lambda$.

One makes an important observation. Besides the usual LI that acts on
the coordinates {\em and} the indices of the field $U^m$, the
Lagrangian (\ref{SUSYaether}) possesses an extra $SO(3,1)$ invariance
that acts {\em only} on the $U^m$ indices. 
In fact, if one treats the index `$m$' as internal, one
  immediately recognizes in (\ref{SUSYaether}) the Lagrangian of 
the standard
  supersymmetric non-linear sigma-model with internal Lorentz
  group. Thus one concludes that at the level of the lower-dimension
  operators the super-aether is equivalent to a Lorentzian sigma-model
  with  K\"ahler potential $f(U^m\bar U_m)$. 
The
symmetries of (\ref{SUSYaether}) contain the product
\[
\text{Lorentz invariance}\times \text{internal}~ SO(3,1)\;.
\] 
Even when the vector $u^m$ acquires non-zero VEV the diagonal subgroup
of this product remains unbroken. This means that the theory for
perturbations around the ground state effectively continues to be
Lorentz invariant. In particular, as we will see explicitly below, all
perturbations have the same dispersion relation. 
Thus LI emerges
as an accidental symmetry due to SUSY even in the Lorentz violating
sector! 
Let us emphasize that this is happening despite that $U_m$ is neutral 
with respect to all (internal) gauge symmetries. Rather, the emergence of LI in 
this sector appears as a consequence of the unit-norm condition.

Let us stress again that this does not mean that LI is completely
restored. Once we go beyond the low-energy theory it is easy to write
higher-dimensional contributions into the super-aether Lagrangian
where $U^m$ is contracted with space-time derivatives, such as the
operators 
\[
\int d^2\theta\; (\d_m U^m)^2+\text{h.c}~,~~~
\int d^2\theta d^2\bar\theta\; \bar{U}^m U^n\,\d_m \bar{U}_l\,\d_n U^l\;
\]
multiplied by the appropriate dimensionful coupling constants. This
removes the accidental internal $SO(3,1)$ symmetry and thus leads to a
genuine breaking of LI by the VEV of $u^m$. Alternatively we can
consider coupling of the super-aether to another sector of the
theory. If this sector contains (at least two) chiral gauge singlets,
it is possible to construct a dimension 4 coupling 
\be
\label{LVdim4U}
{\cal L}^{(4)}_{LV}=\kappa_{(IJ)}\,\int d^2\theta\; 
U^m\Phi_{(I)}\d_m\Phi_{(J)}+\text{h.c.}
\ee
that reduces to the Lorentz violating operator (\ref{LVdim4}) in the
constant aether background. If gauge singlets are absent, as
in the case of MSSM, there are always higher-dimensional couplings
such as, e.g., 
\be
\label{LVdim6U}
{\cal L}_{LV}^{(6)}=
M_{LV}^{-2}\int d^2\theta d^2\bar\theta\;
\bar{U}^m U^n
{\cal D}_m\bar\Phi{\cal D}_n\Phi\;.
\ee 
This is the generalization of (\ref{LVdim6}) to the case of dynamical
aether.
Finally, the internal $SO(3,1)$ will be broken by gravity that
  mixes the space-time and vector indices through covariant derivatives.

It is instructive to write the Lagrangian (\ref{SUSYaether}) in
components. Using (\ref{Umu}) and a similar
decomposition
\be
\label{Lambda}
\Lambda=l(y)+\sqrt{2}\theta\xi(y)+\theta^2H(y)\;,
\ee
after integration over the anticommuting variables we obtain,
\be
\label{SUSYlagr}
{\cal L}={\cal L}_{bos}+{\cal L}_{ferm}\;,
\ee
where the bosonic and fermionic parts are
\begin{align}
\label{boscomp}
{\cal L}_{bos}=M^2_{\text{\ae}}\Big[-&f^{mn}\,\d_r \bar{u}_m\d^r u_n
+f^{mn}\,\bar{G}_m G_n
+\big[H(u_m u^m+1)+2l u_m G^m+\mathrm{h.c.}\big]\Big]\;,\\
\label{fermcomp}
{\cal
  L}_{ferm}=\frac{M^2_{\text{\ae}}}{2}\bigg[&
if^{nr}\,\d_m\bar\eta_n\bar{\sigma}^m \eta_r\notag\\
&\!\!+\!f''\,\bigg(\!\!-i \bar{u}_m\d_r
u_n\;\bar{\eta}^n\bar{\sigma}^r\eta^m
-i \bar{u}_r\d_m u^r\, \bar\eta_n\bar\sigma^m\eta^n
-\,u_m G_n\,\bar\eta^m\bar\eta^n
+\frac{1}{2}(\bar\eta_m\bar\eta_n)(\eta^m\eta^n)\bigg)\notag\\
&\!\!+\!f'''\,\big(-iu_m \bar{u}_n\, \bar{u}_r\d_s u^r\;
\bar{\eta}^m\bar\sigma^s\eta^n
-u_m u_n \bar{u}_s G^s\;\bar\eta^m\bar\eta^n
+u_m \bar{u}_n\,(\bar\eta^m\bar\eta_r)(\eta^n\eta^r)\big)\notag\\
&\!\!+\!\frac{1}{4}f'''' u_m u_n \bar{u}_s \bar{u}_r\,
(\bar\eta^m\bar\eta^n)(\eta^s\eta^r)
-l\eta_m\eta^m-2(\xi\eta_m)u^m+\mathrm{h.c.}\bigg]\;.
\end{align}
Here the ``effective metric'' is
\be
\label{effmetr}
f^{mn}=f'(w)\,\eta^{mn}+f''(w)\,u^m \bar{u}^n\;,
\ee
and primes denote derivatives of the function $f$ with respect to its
argument 
\be
\label{w}
w\equiv u^m\bar{u}_m\;.
\ee

Let us set the fermionic components to zero and concentrate on the
bosonic part (\ref{boscomp}). The equations of motion for the fields
$G^m$, $l$ imply that these fields vanish, i.e. they are
non-dynamical. The field $H$ is a Lagrange multiplier that enforces
the unit-norm constraint on the aether $u^m$. For purely real aether
Eq.~(\ref{boscomp}) reduces to the standard aether Lagrangian
(\ref{Saether}) with 
\be
\label{special}
c_1=2f'(-1)~,~~~c_2+c_3=c_4=0\;.
\ee 
Due to the relation (\ref{equiv}) considerations in flat space-time do
not allow to determine the coefficients $c_2$, $c_3$ separately, but
only their sum. 

\subsection{Vacua and fluctuations}
\label{sec:vacua}

We now classify the ground states of the model in the general case
  of complex aether. 
The unit-norm
constraint implies two equations  
for the real and imaginary parts of
$u^m$,
\bseq
\label{cons*} 
\begin{align}
\label{cons1}
&u_{\mathrm{R}}^m u_{\mathrm{R}\,m}-u_{\mathrm{I}}^m u_{\mathrm{I}\,m}=-1\;,\\
\label{cons2}
&u_{\mathrm{R}}^m u_{\mathrm{I}\,m}=0\;.
\end{align}
\eseq
Consider first the case when the vector
$u^m_\mathrm{R}$ is time-like. By a suitable Lorentz rotation it can
be aligned with the time axis. Then due to (\ref{cons2})
$u^m_\mathrm{I}$ has only spatial components and by a 3d rotation we
direct it along the third axis. Taking into account (\ref{cons1}) we
obtain a family of ground states,
\be
\label{ubackgr}
u^m_{vac}=(\cos\a,\, 0,\,0,\, i\sin\a)\;,
\ee   
parameterized by $\a\in[0,\pi/2]$. Note that for $\a\neq 0$ the ground
state, besides breaking boosts, breaks also the spatial
isotropy. Indeed, for $\a\neq 0,\,\pi/2$ the group of invariance of
(\ref{ubackgr}) reduces to the $SO(2)$ rotations in the
(1,2)-plane. For $\a=\pi/2$ the unbroken group is enhanced up to
$SO(2,1)$. Only for $\a=0$ one recovers the 3-dimensional $SO(3)$
rotations. Thus we conclude that SUSY has introduced a qualitatively
new feature into the aether model: existence of a family of
inequivalent vacua, generally breaking spatial isotropy together with
LI. Let us stress, however, that breaking of rotational invariance
has a very different physical meaning from the breaking of boosts. In
a sense, the boost invariance is broken {\em explicitly}. Indeed,
there is no state in the theory where it would be
recovered\footnote{One may think that the boost invariance can be at
  least partially restored, as it appears to happen in the vacuum
  (\ref{ubackgr}) at $\a=\pi/2$ and in the vacua
  (\ref{ubackgr1}). However, we are going to see that these vacua
  contain a ghost in the spectrum of perturbations and thus are
  dynamically inaccessible.}. On the
other hand, breaking of the rotation group is truly {\em spontaneous}:
the symmetry breaking configurations smoothly connect to the vacuum
with $\a=0$ where the full $SO(3)$ group is restored.

Apart from (\ref{ubackgr}) the constraints (\ref{cons*}) possess a
branch of solutions when both $u^m_\mathrm{R}$ and $u^m_\mathrm{I}$
are space-like. Due to the orthogonality condition (\ref{cons2}) we
can direct these vectors along the second and third axes. Then
Eq.~(\ref{cons1}) yields
\be
\label{ubackgr1}
u^m_{vac}=(0,0,\sh\b,i\ch\b)\;,
\ee 
where $\b>0$. These vacua leave unbroken the subgroup $SO(1,1)$ of the
Lorentz group corresponding to the boosts in the (0,1)-plane.

To get more insight into the dynamics let us study small perturbations
around the vacua. We start with the background
(\ref{ubackgr}). Writing 
\be
\label{perturb}
u^m=u^m_{vac}+v^m\;,
\ee
one finds that the unit-norm constraint relates the components $v_0$
and $v_3$. It is convenient to express them in terms of a single field
$v_\|$,
\[
v_0=-iv_\|\sin\a~,~~~~v_3=v_\|\cos\a\;.
\] 
Substituting this into (\ref{boscomp}) we obtain the quadratic
Lagrangian, 
\be
\label{quadrL}
{\cal L}_{bos}=-M^2_{\text{\ae}}
\Big[\big(-wf'+(1-w^2)f''\big)\,\d_m v_\|\, \d^m\bar{v}_\|
+f' \sum_{a=1,2}
\d_m v_a\, \d^m \bar{v}_a\Big]\;,
\ee
where 
\be
\label{w1}
w=-\cos 2\a\;. 
\ee
Note that this Lagrangian is effectively Lorentz invariant if we treat
the perturbation components $v_1$, $v_2$, $v_\|$ as scalars. In
particular, all modes have the same linear dispersion relation with
unit velocity. This is the manifestation of the emergent LI discussed
above. On the other hand, the symmetry between $v_\|$ and $v_1$, $v_2$
is clearly broken in (\ref{quadrL}). The Lagrangian is free of
ghosts or any other pathologies as long as 
\[
f'>0~,~~~~-wf'+(1-w^2)f''>0\;.
\]
Therefore, the model allows breaking of spatial isotropy (in the
  vacuum with $\a\neq 0$) and still possesses completely stable
  spectrum of perturbations. Note that this situation cannot be
  realized in the real-aether model of 
 Sec.~\ref{aether}. In that case considering space-like aether
 corresponds to switching the sign in the constraint
 (\ref{norm}). Then it is straightforward to check that the spectrum
 of perturbations contains a ghost. Instead, for the super-aether
 model 
changing the sign in the
constraint (\ref{Uconstr}) does not affect the physics: it simply
corresponds to multiplying $U^m$ by $i$. We have seen that the model
is stable 
so long as one of the components of the vector -- 
its real or imaginary part -- is
time-like.

In the purely spatial background (\ref{ubackgr1}) the situation
is different and, actually, is analogous to the case of spatial real
aether. Here
the unit-norm constraint relates $v_2$ and $v_3$, and the quadratic
Lagrangian reads
\be
\label{quadrL1}
{\cal L}_{bos}=-M^2_{\text{\ae}}
\Big[-f'\,\d_m v_0\,\d^m\bar{v}_0+f'\,\d_m v_1\,\d^m\bar{v}_1
+\big(wf'+(w^2-1)f''\big)\,\d_m v_\|\, \d^m\bar{v}_\|
\Big]\;,
\ee
where now $w=\ch 2\b$ and $v_\|$ is introduced as
\[
v_2=-iv_\|\ch\b~,~~~~v_3=v_\|\sh\b\;.
\]
Clearly, one of the modes is always a ghost. Thus the
purely space-like aether background is pathological and we do not
consider it in what follows.

Finally we analyze the fermionic content of the model. 
The spinor field $\xi$ in the last line of
(\ref{fermcomp}) is a Lagrange multiplier that enforces
the constraint 
\be
\label{consferm}
u^m\eta_{m\gamma}=0\;.
\ee
Solving this constraint in the
background (\ref{ubackgr}),
\[
\eta_{0\gamma}=-i\eta_{\|\gamma}\sin\a~,~~~~
\eta_{3\gamma}=\eta_{\|\gamma}\cos\a\;,
\]
we obtain
\be
\label{quadrLF}
{\cal L}_{ferm}=\frac{iM^2_\text{\ae}}{2}\Big[
\big(-wf'+(1-w^2)f''\big)
\d_m\bar\eta_\|\bar\sigma^m\eta_\|
+f'\sum_{a=1,2}
\d_m\bar\eta_a\bar\sigma^m\eta_a
+\mathrm{h.c.}\Big]+\ldots\;,
\ee
where $w$ is given by (\ref{w1}) and dots stand for the terms with
higher powers of the fermionic fields. This Lagrangian describes three
fermions in the Weyl representation 
with respect to the emergent LI that match the three
complex scalars of (\ref{quadrL}).

\section{Breaking of SUSY}
\label{breaking}

In this section we discuss the effects of SUSY breaking on the
super-aether model. The breaking of SUSY is conveniently described
using spurion superfields whose higher components have non-vanishing
VEVs. We will assume the spurions and their non-zero components to be
Lorentz scalars, so that they do not source additional violation of
LI.  
Then the SUSY breaking Lagrangian of the
lowest dimension is,
\be
\label{SUSYbreak}
{\cal L}_{SB}=-M^2_{\text{\ae}}\int d^2\theta d^2\bar\theta\,
\big[S_{(1)}\, g_{(1)}(U^m \bar{U}_m)+S_{(2)}\,g_{(2)}(U^m \bar{U}_m)\big]\;,
\ee
where $g_{(1)}$, $g_{(2)}$ are arbitrary functions and the spurions have
the form
\bseq
\label{spurions}
\begin{align}
\label{spurion1}
&S_{(1)}=m_{(1)}^2\,\theta^2\bar\theta^2\;,\\
\label{spurion2}
&S_{(2)}=m_{(2)}(\theta^2+\bar\theta^2)\;.
\end{align}
\eseq
The parameters $m_{(1)}$, $m_{(2)}$ have the dimension of mass; they
set the scale of SUSY breaking in the aether sector.
The Lagrangian (\ref{SUSYbreak}) reads in components
\be
\label{SBcomp}
{\cal L}_{SB}=-M^2_{\text{\ae}}\bigg[m_{(1)}^2g_{(1)}(w)
+m_{(2)}\Big(g_{(2)}'(w)\bar{G}_m u^m
-\frac{1}{2}\,g_{(2)}''(w)u_m u_n(\bar\eta^m\bar\eta^n)
+\mathrm{h.c.}\Big)\bigg]\;,
\ee
where $w$ is defined in (\ref{w}). This must be combined with the
supersymmetric part
(\ref{boscomp}), (\ref{fermcomp}).

It is straightforward to work out the effect of the first term in
(\ref{SBcomp}). Consider the aether vacuum (\ref{ubackgr}). One
observes that the modulus $\a$ acquires a potential proportional to
$g_{(1)}(-\cos 2\a)$. If $g_{(1)}'(-1)>0$ the modulus is stabilized at
$\a=0$. Thus we conclude that SUSY breaking lifts the degeneracy of
the aether vacua and, under broad assumptions, singles out the purely
time-like aether configuration. Note that this configuration is real
and preserves the spatial isotropy. The imaginary part of
the aether perturbations acquires a mass of order
$m_{(1)}$ in this vacuum. 

The effect of the contribution proportional to $m_{(2)}$ in
(\ref{SBcomp}) is subtler. At first sight, the second term in the round
brackets seems to be a candidate for the fermionic mass term. However,
due to the constraint (\ref{consferm}), this term actually vanishes to
quadratic order in the purely real aether background. 
Nevertheless, the fermions do get masses, though in an indirect
way. We observe that the $m_{(2)}$-contribution modifies the equations
for the non-dynamical fields $G^m$, $l$, so that the field $l$ no
longer vanishes. Instead we obtain in the vacuum  
\be
\label{lambdavac}
l_{vac}=\frac{m_{(2)}g_{(2)}'}{2}\;.
\ee
Substituted into the last line of the fermionic Lagrangian 
(\ref{fermcomp}) this endows $\eta^m$ with the mass of order $m_{(2)}$.

To sum up, we have found that the general result of SUSY breaking is
to generate masses for the fermions and the imaginary part of the
aether. Below the SUSY breaking scale one is left with the ordinary
real aether described in Sec.~\ref{aether}. However, the memory of the
SUSY origin of the theory is not totally lost: it is encoded in the
relations (\ref{special}) between the coefficients of the aether
action\footnote{Strictly speaking, the SUSY breaking will also affect
  the relations (\ref{special}) through higher-dimensional
  operators. However, these corrections are suppressed by powers of
  the ratio of the SUSY breaking scale to $M_{\text{\ae}}$. They are
small provided there is a hierarchy between these scales.}.

\section{Discussion and outlook}
\label{future}

In this paper we have considered a prototypical model of Lorentz
violation with a dynamical preferred frame -- the Einstein-aether
theory -- and have constructed the supersymmetric extension of the
aether sector. Our model allows to implement the SUSY-based mechanism
of \cite{GrootNibbelink:2004za,Bolokhov:2005cj} that ensures emergence
of LI at low energies within the SM sector. SUSY turned out to be more
powerful in this respect than one could a priori expect: it leads to
emergent low-energy LI even within the aether sector itself. We have
found that the dynamics of the super-aether is richer than of its
non-SUSY counterpart. In particular, the model possesses a family of
inequivalent vacua exhibiting different symmetry breaking
patterns. Remarkably, the model tolerates a certain breaking of
spatial isotropy while remaining stable and ghost free. 
Finally, we have analyzed the effects of SUSY breaking.

In this paper we have restricted to the case of global SUSY. An
important next step is the coupling of the super-aether model to
supergravity. This task appears straightforward but presents a 
non-trivial technical exercise. 
Leaving it for the future, let us mention the following subtlety.
It is impossible to embed the model of the
present paper in the framework of the minimal
SUGRA~\cite{Stelle:1978ye,Ferrara:1978em}. 
The reason is that in this framework, due to non-vanishing
supercurvature, the covariant spinor derivatives do not anticommute
when acting on a vector superfield, 
\be
\{\bar{\nabla}_{\dot\a},\bar{\nabla}_{\dot\b}\}U^m\neq 0\;.
\ee
This prevents the definition of the chiral aether vector
\cite{Fischler:1979ty}. To overcome this obstruction one has to resort
to the non-minimal formulation of 
SUGRA~\cite{Breitenlohner:1976nv,Siegel:1978mj} where 
it is possible to construct improved spinor
derivatives, that do anticommute~\cite{Brown:1979xt}. 

It would be interesting to generalize the construction of this paper to
other theories with dynamical preferred frame. To illustrate possible
problems let us consider the khrono-metric model of
\cite{Blas:2010hb}. This model arises as the low-energy limit of the
healthy Ho\v rava gravity \cite{Horava:2009uw,Blas:2009qj}
and is closely related to the Einstein-aether theory
\cite{Jacobson:2010mx}. Violation of LI in this model is described by
a {\em real} scalar field $\varphi$ -- the khronon -- that has non-vanishing
time-like gradient. One imposes invariance under reparameterizations
of $\varphi$,
\be
\label{reparam}
\varphi\mapsto\tilde\varphi(\varphi)\;,
\ee
which implies that $\varphi$ enters the Lagrangian only through the
unit time-like vector  
\be
\label{khronou}
u_m=\frac{\d_m\varphi}{\sqrt{-\d_n\varphi\d^n\varphi}}\;.
\ee
In terms of this vector the action has the same form (\ref{Saether})
as the aether action\footnote{For the hypersurface-orthogonal vector
  (\ref{khronou}) the four terms in the aether action are not
  independent. One of them can be eliminated in favor of the
  rest.}. Note, however, that in terms of the fundamental field
$\varphi$ the action contains higher derivatives leading to
fourth-order equations of motion. As discussed in \cite{Blas:2010hb},
this does not lead to inconsistencies. Due to the special structure of
the fourth-order operator it is always possible to choose a coordinate
frame in such a way that the equations contain only two
time-derivatives, the extra two derivatives being purely
space-like. The preferred frame is determined by the khronon field
itself. Namely, the time coordinate must be chosen to coincide with
the background value of the field,
\be
\label{khronogauge}
t=\varphi_{\mathrm{background}}\;.
\ee
We stress that the possibility to make this choice is crucial for the
consistency of the model. As we now discuss, this property is lost
when we attempt to construct a SUSY extension of the model, and higher
time derivatives persist in the equations of motion. 

By the analogy with the aether case, a natural choice of the khronon
superfield is a chiral scalar,
\be
\label{Phi}
\Phi=\varphi(y)+\sqrt{2}\theta^\a\psi_\a(y)+\theta^2 F(y)\;.
\ee 
Importantly, this requires that the khronon field $\varphi$ be
complexified. From this scalar one constructs a vector superfield
\be
\label{khronoU}
U_m=\frac{\d_m\Phi}{\sqrt{-\d_n\Phi\d^n\Phi}}\;.
\ee
The lowest components of $U_m$ and $\Phi$ are related by
Eq.~(\ref{khronou}). 
Note that $U_m$ is automatically chiral and satisfies the unit-norm
constraint (\ref{Uconstr}). 
The Lagrangian of the theory has the form (\ref{SUSYaether}) with the
only 
difference that now there is no need for a Lagrange multiplier. 

The khronon vacuum corresponding to (\ref{ubackgr}) has the form
\be
\label{khronovac}
\varphi_{vac}=t\cos\alpha +iz\sin\a\;.
\ee
Note that this vacuum preserves SUSY despite of the fact that the SUSY
variation of the fermionic component is non-zero, 
\be
\label{psivar}
\delta\psi=i\sqrt{2}\sigma^m\bar\zeta\,\d_m\varphi_{vac}
=\sqrt{2}(i\sigma^0\cos\a-\sigma^3\sin\a)\bar\zeta\;,
\ee
where $\zeta$ is the parameter of the SUSY transformation. 
The reason is the large symmetry group of the theory. The expression
(\ref{khronoU}) 
is invariant not only under arbitrary reparameterizations of $\Phi$,
but also under all transformations of the form 
\be
\Phi\mapsto\tilde\Phi(\Phi,\theta^\a)
\ee
that involve arbitrary dependence on the holomorphic Grassmann
coordinates. An appropriate transformation of this form compensates
the variation (\ref{psivar}). 

Consider now small perturbations on top of the vacuum
(\ref{khronovac}). 
For our argument it is sufficient to concentrate on the perturbations
of the khronon, leaving aside the fermionic and auxiliary fields.
We write 
\be
\label{khronoflu}
\varphi=\varphi_{vac}+\chi\;,
\ee
and expand the action to quadratic order in $\chi$.
We can use the result (\ref{quadrL}) for the aether:
it remains to express the aether perturbations $v_{1,2}$, $v_\|$ in terms
of $\chi$. A straightforward calculation yields
\bseq
\label{vis}
\begin{align}
&v_a=\d_a\chi~,~~~~a=1,2\;,\\
\label{v3}
&v_\|=\cos\a\,\d_3\chi-i\sin\a\,\dot\chi\;.
\end{align}
\eseq
If $\a=0$ we recover the quadratic Lagrangian of \cite{Blas:2010hb}
for the (complexified) khronon perturbations containing only first
time derivatives of $\chi$. However, in the general case $\a\neq 0$
the action involves second time derivatives $\ddot\chi$ and results
in the fourth-order equations of motion. This leads to very fast
instabilities that are clearly unacceptable. 
Note that the problem cannot be eliminated
even by breaking SUSY or simply by choosing 
the background with $\alpha=0$,
since the higher time derivatives will inevitably appear in interaction terms.
The origin of this
behavior can be traced back to the fact that the khronon is complex and therefore
there is no real coordinate system in which it coincides with the would-be preferred 
time.
Thus we conclude that new
ideas are required to supersymmetrize the khronon model and,
eventually, its UV completion -- Ho\v rava gravity.

Another issue related to the problem of constructing a SUSY Ho\v rava
gravity, but also interesting on its own right, is the extension of
the notion of SUSY to theories invariant under anisotropic scaling of
space and time coordinates. Recently, it has been demonstrated 
\cite{Redigolo:2011bv} that the requirements of the anisotropic scaling
and gauge invariance are incompatible with the non-relativistic
SUSY algebra (\ref{3dSUSY}). Namely, it is impossible to construct any
gauge theory\footnote{Still, there are some non-trivial theories
  without gauge invariance that realize both the anisotropic scaling
  and (\ref{3dSUSY}) \cite{Redigolo:2011bv}.} 
with anisotropic scaling invariant under (\ref{3dSUSY}). As a way out
one could try to deform the SUSY algebra by allowing non-linear
functions of the spatial momentum $P_a$ appear on the r.h.s. of the
commutation relations \cite{Xue:2010ih}. This would make SUSY
compatible with the anisotropic scaling. However, as pointed out in
\cite{Redigolo:2011bv}, construction of interacting models invariant
under the deformed algebra may present a non-trivial task because 
the SUSY transformations now contain higher spatial
derivatives. In particular, one cannot use the superfield formalism to
build invariant Lagrangians. Indeed, 
 any attempt to realize the supercharges on
the superspace will involve higher-order differential operators. These
supercharges will not satisfy the Leibniz rule,
implying that the product of two superfields does not transform
as a superfield and thus the superfield formalism becomes useless.
At present only free theories
invariant under the deformed SUSY have been constructed 
\cite{Redigolo:2011bv}.

\paragraph*{Acknowledgments}

We are grateful to Diego Blas, Sergei Demidov,
Dmitry Gorbunov, Jean-Luc Lehners, Riccardo Rattazzi, 
Thomas Sotiriou and Giovanni Villadoro for useful
discussions. S.S. thanks the Theory Division of IFAE,
Barcelona, for the warm hospitality during his visit. 
This work was supported in part by 
the Russian Ministry of Education and Science under the state
contract 02.740.11.0244~(S.S.),
the Grants of the President of Russian Federation
NS-5525.2010.2 and MK-3344.2011.2~(S.S.) and the RFBR grants
11-02-92108, 11-02-01528~(S.S.).

\end{document}